\documentclass[journal]{IEEEtran}

\makeatletter

\newcommand{\Rmnum}[1]{\expandafter\@slowromancap\romannumeral #1@}
\makeatother

\usepackage{amsmath}

\usepackage{graphicx}
\usepackage{epstopdf}
\usepackage{subfigure}

\usepackage[justification=centering]{caption}

\usepackage{setspace}

\usepackage{enumerate}

\usepackage{amsfonts}

\usepackage{booktabs}

\usepackage{color}

\begin{document}

% ====paper title===========
\title{ELM-based Frame Synchronization in Burst-Mode Communication Systems with Nonlinear Distortion}

%\author{Author 1, Author 2, Author 3,Author 4 and Author 5}
\author{Chaojin~Qing,~\IEEEmembership{Member,~IEEE,}
        Wang~Yu, Bin Cai,
        ~Jiafan~Wang
        and~Chuan~Huang,~\IEEEmembership{Member,~IEEE} %<-this  stops a space

%\thanks{Manuscript received xx, 2019; accepted xx, 2019. Date of publication xxx, 2019; date of current version xxx, 2019.
\thanks{This work is supported in part by the National Key Research and Development Program (Grant 2018YFB1800800), the Key Projects of Education Department of Sichuan Province (Grant 15ZA0134), and the Major Special Funds of Science and Technology of Sichuan Science and Technology Plan Project (Grant 19ZDZX0016) of China.}
\thanks{Chaojin Qing, Wang Yu and Bin Cai are with School of Electrical Engineering and Electronic Information, Xihua University, Chengdu, China. (e-mail: qingchj@uestc.edu.cn)}% <-this % stops a space
\thanks{Jiafan Wang is with Synopsys Inc., 2025 NE Cornelius Pass Rd, Hillsboro, OR 97124 USA. (email: jifanw@gmail.com)}% <-this % stops a space
\thanks{Chuan Huang is with the National Key Laboratory of Science and Technology on Communications, University of Electronic Science and Technology of China, Chengdu, China (e-mail:  huangch@uestc.edu.cn).}}

% The paper headers
\markboth{IEEE WIRELESS COMMUNICATIONS LETTERS,~Vol.~XX, No.~XX, XXX~2020}%
{Qing \MakeLowercase{\textit{et al.}}: ELM-BASED FRAME SYNCHRONIZATION IN BURST-MODE COMMUNICATION SYSTEMS WITH NONLINEAR DISTORTION}

% make the title area
\maketitle

%===========摘要==============
\begin{abstract}
 In burst-mode communication systems, the quality of frame synchronization (FS) at receivers significantly impacts the overall system performance. To guarantee FS, an extreme learning machine (ELM)-based synchronization method is proposed to overcome the nonlinear distortion caused by nonlinear devices or blocks. In the proposed method, a preprocessing is first performed to capture the coarse features of synchronization metric (SM) by using empirical knowledge. Then, an ELM-based FS network is employed to reduce system's nonlinear distortion and improve SMs. Experimental results indicate that, compared with existing methods, our approach could significantly reduce the error probability of FS while improve the performance in terms of robustness and generalization.

 %is robust and generic with the change of system parameter.
\end{abstract}

%========= 关键字============
\begin{IEEEkeywords}
frame synchronization, extreme learning machine, non-linear distortion, synchronization metric.
\end{IEEEkeywords}

\IEEEpeerreviewmaketitle

%=========第一章introduction （空一行表示分段）============
\section{Introduction}

\IEEEPARstart{N}{owadays} the burst-mode transmission is pervasively applied in modern communication systems, such as Internet of Things (IoT) \cite{c1}, wireless local area networks (WLAN) \cite{c2}, etc. In burst-mode communication system (BCS), frame synchronization (FS) is the foundation of the overall system performance, and is always assumed to be obtained at the receiver. However, the BCS has a large number of non-linear devices or blocks, e.g., high power amplifier (HPA), digital to analog converter (DAC), etc., inevitably causing nonlinear distortion \cite{c5}, \cite{c3}. Usually, synchronization precedes channel estimation, signal demodulation, etc., and thus first encounters these nonlinear distortions, degrading receiver's FS performance (e.g., the error probability performance). Owing to the lack of considerations for nonlinear distortion, the existing methods (e.g., correlation-based FS \cite{c6}, etc) are facing great challenges.

In recent years, machine learning has drawn considerable attention due to its prominent ability to cope with nonlinear distortion \cite{c8}, \cite{c9}. The machine learning, in particular deep learning (DL) has been applied in wireless communication, e.g., signal detection \cite{c9}, precoding \cite{c7}, channel state information (CSI) feedback \cite{c10}, channel estimation \cite{c11}, \cite{c4} etc. Yet, very limited works are focused on DL-based FS. One related work \cite{c12} investigated the DL-based timing synchronization, yet shows a higher timing error probability than conventional matched filtering. In addition, these DL-based approaches suffer from many difficulties such as complex parameter tuning, and long-time training \cite{c10}, etc.

Unlike the DL-based approaches, the extreme learning machine (ELM) is a single-hidden layer feed-forward neural network, i.e., the gradient back-propagation (BP) is not required, possessing many advantages, e.g., randomly generating for input weight and hidden bias, fast learning speed (hundreds of times faster than that of BP algorithm), and good generalization performance, etc., \cite{c13}, \cite{c14}. Inspired by these advantages, an ELM-based FS is proposed in this paper to improve the training sequence-based method, e.g., correlation-based FS \cite{c6}. Due to the loss of training sequence's orthogonality, the training sequence-based FS is difficult to apply in the scenario of nonlinear distortion. In the proposed method, a preprocessing is first performed to coarsely capture the features of synchronization metric (SM) by using empirical knowledge. Then, an ELM network is employed to alleviate system's nonlinear distortion and improve SMs. Compared with the correlation-based FS \cite{c6} and recent FS method in \cite{c15}, the proposed method can effectively reduce the error probability of FS for the cases with nonlinear distortion. Furthermore, with the parameter impacts, the proposed method shows a stable improvement given the change of system parameters.

The remainder of this paper is structured as follows: In Section \Rmnum{2}, we briefly describe the system model. In Section \Rmnum{3}, the ELM-based FS method is specifically presented, and the numerical simulation and analysis are given in Section \Rmnum{4}, the Section \Rmnum{5}  concludes our work.

Notations: Bold lowercase and uppercase letters denote vectors and matrices respectively; italicized letters denote variables;  $(\cdot)^{T}$, $(\cdot)^{H}$, $(\cdot)^{-1}$ and ${\left( \cdot\right)^\dag }$ denote the transpose, conjugate transpose, matrix inversion, Moore–Penrose pseudoinverse, respectively; ${\bf{0}}_{N}$ is ${N} \times 1$ vector with ${N}$ zero elements; ${{\left\| \cdot  \right\|}_{2}}$  is the Frobenius norm; $\left| x \right|$  denotes the absolute value of $x$ and $\left| \mathbf{x} \right|$  denotes the absolute value operation to the every elements of vector $\mathbf{x}$.

%==========第二章System model===============
\section{SYSTEM MODEL}

%=============正文===================
Considering a frame-based BCS, the transmitted frame format is illustrated in Fig. 1(a), which consists of $N_s$ training symbols ${\bf{s}} = {\left[ {{s_1}, {s_2},\cdots ,{s_{{N_s}}}} \right]^T}\in\mathbb{C}^{{N_s}\times1}$, ${N_g}$ empty symbols ${\mathbf{0}^T_{N_g}}$, and $N_d$ data symbols ${\bf{d}} = {\left[ {{d_1},{d_2},\cdots,{d_{{N_d}}}} \right]^T}\in\mathbb{C}^{{N_d}\times1}$. To guarantee the training symbol and data symbol are allocated the same transmitted power, $E \{|s_i|^2 \} = E \{|d_j|^2 \} = P$, $i = 1,2,\cdots,N_s$, $j=1,2,\cdots,N_d$, is considered in this paper. The transmitted frame ${\bf{x}}\in \mathbb{C} ^ {M \times 1}$ is formed by ${\left[ {{\bf{s}}^T, {\mathbf{0}^T_{N_g}},{\bf{d}}^T} \right]^T} $, where $M$ is the frame length. Similar to \cite{c15}, ${N_g}$ empty symbols are employed to mitigate the multi-path channel dispersion. Fig. 1(b) presents the system model, in which the nonlinear distortion (due to the existence of nonlinear blocks or devices, such as HPA, DAC, etc \cite{c5}) is encountered by the frame ${\bf{x}}$, and then the distorted signals are transmitted. At the receiver, the observation of transmitted training sequence ${\bf{s}}$, denoted as ${\bf{y}} \in \mathbb{C}^{M \times 1}$, can be expressed as follows \cite{c15}

\begin{equation}
\label{eq:1}
{\bf{y}} = {\bf{\widetilde S\widetilde h}} + {\bf{n}},
\end{equation}
where $\mathbf{n} \in \mathbb{C}^{M \times 1}$  is the complex additive white Gaussian noise (AWGN) vector whose entries are with zero-mean and variance $\sigma ^2$. The $M \times N$ complex matrix $\mathbf{\widetilde{S}}$, which consists of the distorted and shifted version of transmitted frame $\bf{x}$, could be defined as
\begin{equation}
{\bf{\widetilde S}} = \left[ \begin{array}{l}
{{\widetilde s}_1} \qquad {\rm{       0       }} \qquad \cdots \\
\  {\rm{ }} \vdots \qquad  {\rm{       }}{{\widetilde s}_1} \qquad {\rm{      }} \ddots \\
{{\widetilde s}_{{N_s}}} \quad \ \, {\rm{   }} \vdots \qquad \, {\rm{       }} \ddots \\
\; {\rm{ 0     }} \quad \  \ {{\widetilde s}_{{N_s}}} \quad \ {\rm{    }} \ddots \\
\ {\rm{ }} \vdots \qquad \  {\rm{       0       }} \qquad \ddots \\
\qquad \quad {\rm{          }} \vdots \qquad \,{\rm{       }} \ddots
\end{array} \right],
\end{equation}
where ${\widetilde s_i},i = 1,2, \cdots ,{N_s}$ stands for the variant of the training symbol ${s_i}$ due to the nonlinear distortion, and $N$ is the size of search window. In equation (1), $\mathbf{\widetilde h} \in \mathbb{C}^{N \times 1}$ represents the extended vector of channel impulse response (CIR), which can be written as
\begin{equation}\label{eq:3}
{\bf{\widetilde h}}{\rm{ = }}{\left[ {\underbrace {0,\cdots,0}_\tau,{{\bf{h}}^T},\underbrace {{\rm{0}}, \cdots, {\rm{0}}}_{N - L - \tau }} \right]^T},
\end{equation}
where $\tau $ is the frame boundary offset to be estimated with $0 \le \tau  \leq M - {N_s}-1$. In (\ref{eq:3}), ${\bf{h}} = {\left[ {{h_1},{h_2},\cdots ,{h_L}} \right]^T}$ denotes the finite CIR vector of $L$ samples memory, where ${h_l},l = 1,2, \cdots ,L$ represents the complex-valued CIR of the $l$th path.

%==========插入图1================
\begin{figure}
\includegraphics[width=0.4\textwidth]{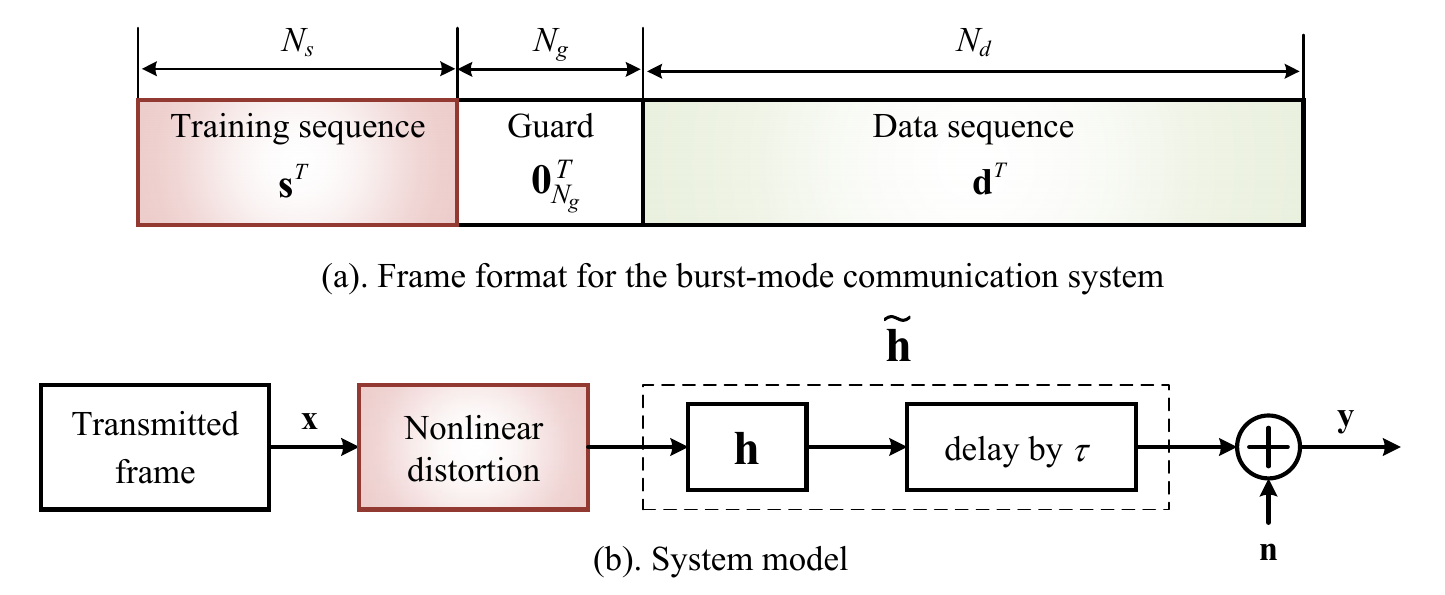}
\caption{Frame format and system model}
\label{fig1}
\end{figure}

%==========第三章 ELM帧同步==========
\section{ELM FRAME SYNCHRONIZATION}
In this section, a preprocessing for FS is first described in section III-A, followed by an ELM network (given in section III-B). According to \cite{c12}, the error probability of DL-based timing synchronization is far higher than that of matched filtering, while similar behaviors are also observed in FS experiments where the ELM-based networks are employed for the scenarios with or without nonlinear distortion. Thus, a preprocessing is employed to capture the coarse features of SM.

%========3.1 帧同步预处理==================
\subsection{Preprocessing of Frame Synchronization}
In the conventional approaches, $\tau $ can be estimated by using cross-correlation based SM \cite{c6}, \cite{c15}, i.e.,
\begin{equation}\label{eq:4}
{\widehat{\tau}_{\mathrm{cross-correlation}}} = \mathop {\arg \max }\limits_{0 \le t \le M - {N_s}-1} {\left| {{{\bf{s}}^H}{{\bf{y}}_{t:t + {N_s} - 1}}} \right|^2},
\end{equation}
where ${{\bf{y}}_{t:t + {N_s} - 1}}$ represents the elements of ${\bf{y}}$ from $t$ to $t+{N_s}-1$. In this paper, the existing methods for computing SMs, e.g., cross-correlation based method in \cite{c6}, are viewed as \emph{empirical knowledge}. It should be noted that besides the cross-correlation based SM, other SMs could also be applied in our method with the similar processing. Denoting the cross-correlation based SM as
\begin{equation}\label{eq:5}
{\mathbf{\Gamma}_t} = {\left| {{{\bf{s}}^H}{{\bf{y}}_{t:t + {N_s} - 1}}} \right|^2}, t = 0,1,\cdots,M - {N_s}-1,
\end{equation}
then the SM vector $\mathbf{g} \in \mathbb{R} {^{(M - {N_s})\times 1}}$ can be given by
\begin{equation}\label{eq:6}
{\mathbf{g}} = {\left[ {{\mathbf{\Gamma} _0},{\mathbf{\Gamma} _1}, \cdots ,{\mathbf{\Gamma} _{M - {N_s}-1}}} \right]^T}.
\end{equation}
For easy operation of ELM network, we normalize $\mathbf{g}$ in (\ref{eq:6}) as
\begin{equation}\label{eq:7}
{\bf{\bar g }} = {{\bf{g }} \mathord{\left/
 {\vphantom {{\bf{g }} {{{\left\| {\bf{g }} \right\|}_2}}}} \right.
 \kern-\nulldelimiterspace} {{{\left\| {\bf{g }} \right\|}_2}}}.
\end{equation}
Then, ${\bf{\bar g }}$ is used as the input of ELM network. We will employ ELM network for FS to decrease nonlinear distortion and improve SMs, which is elaborated in the following subsection.

%=========3.2 基于ELM的帧同步=========
\subsection{ELM-based Frame Synchronization }
The ELM-based FS includes offline and online procedures, which are elaborated in Table I and Table II, respectively.

%==========画table1===============
\begin{table}
	\centering
	\caption{Offline training procedure}
	\label{tab:table1}
	\centering
	\begin{tabular}{p{0.95\linewidth}}
	\toprule
\\
~~Given a training set $\left\{ {\left. {\left( {{{{\bf{\bar g }}}_i},{{\bf{T}}_i}} \right)} \right|i = 1,2, \cdots ,{N_t}} \right\}$, hidden neuron number $\widetilde{N}$ and real-valued activation function $\sigma \left(  \cdot  \right)$, the training steps are summarized as follows:
	\begin{enumerate}[step 1 :]
		\item Randomly choose the input weight ${\bf{W}} \in \mathbb{R}{^{\widetilde{N} \times \left( {M - {N_s}} \right)}}$ and the  hidden bias ${\bf{b}} \in \mathbb{R} {^{\widetilde{N} \times 1}}$.
		\item According to ${\bf{W}}$ and ${\bf{b}}$, calculate the hidden layer output ${{\bf{H}}_i} \in \mathbb{R}{^{\widetilde{N} \times 1}}$ and construct a training output matrix ${\bf{H}} \in \mathbb{R} {^{\widetilde{N} \times {N_t}}}$ as given in (10).
		\item Use the desired label ${{\bf{T}}_i}$ to construct a label matrix ${{\bf{T}}}$ according to (11), and then compute output weight ${\bf{\Upsilon}} \in \mathbb{R} {^{\left( {M - {N_s}} \right) \times \widetilde N}}$ according to (12).
	\end{enumerate}
	\\ \bottomrule
	\end{tabular}
\end{table}

For offline training, ${N_t}$ samples, i.e., $\{ \left( {{{{\bf{\bar g }}}_i},{{\bf{T}}_i}} \right) \}$, $i = 1,2, \cdots ,{N_t}$ , are collected to form a  training set, where ${{\bf{T}}_i}$ is the offset label of the $i$th sample. Denoting the offset of the $i$th sample as ${\tau ^{\left( i \right)}}$, the label ${{\bf{T}}_i}$ can be encoded according to one-hot mode, i.e.,
\begin{equation}\label{eq:8}
{{\bf{T}}_i} = {\left[ {\underbrace {0, \cdots, 0}_{{\tau ^{\left( i \right)}}}{\rm{    ,1,    }}\underbrace {{\rm{0,}} \cdots, 0}_{M - {N_s} - {\tau ^{\left( i \right)}} - 1}} \right]^T}.
\end{equation}
As shown in Table I, during the offline training procedure, the input weight ${\bf{W}} \in \mathbb{R}{^{\widetilde N \times \left( {M - {N_s}} \right)}}$ and hidden bias ${\bf{b}} \in \mathbb{R} {^{\widetilde N \times 1}}$ of ELM network are randomly chosen, where $\widetilde N$ is the hidden neuron number. Then, the hidden layer output ${{\bf{H}}_i}\in \mathbb{R} {^{\widetilde N \times 1}}$ can be given by
\begin{equation}\label{eq:9}
{{\bf{H}}_i} = \sigma \left( {{\bf{W}}{{{\bf{\bar g }}}_i} + {\bf{b}}} \right),
\end{equation}
where $\sigma \left(  \cdot  \right)$ denotes the activation function such as sigmoid, hyperbolic tangent (tanh) , rectified linear units (ReLU) \cite{c16}, etc. By collecting ${{\bf{H}}_i}$, a training output matrix ${\bf{H}} \in {^{\widetilde N \times {N_t}}}$ can be constructed as
\begin{equation}\label{eq:10}
 {\bf{H}} = \left[ {{{\bf{H}}_1},{{\bf{H}}_2}, \cdots ,{{\bf{H}}_{{N_t}}}} \right].
\end{equation}
From the training set $\left\{ {\left. {\left( {{{{\bf{\bar g }}}_i},{{\bf{T}}_i}} \right)} \right .} \right\}$, the training labels $\{ {{\bf{T}}_i} \}$ can be used to form a label matrix ${\bf{T}} \in \mathbb{R}{^{\left( {M - {N_s}} \right) \times {N_t}}}$, i.e.,
\begin{equation}\label{eq:11}
{\bf{T}} = \left[ {{{\bf{T}}_1},{{\bf{T}}_2}, \cdots ,{{\bf{T}}_{{N_t}}}} \right].
\end{equation}
According ${\bf{H}}$ and ${\bf{T}}$, the output weight ${\bf{\Upsilon }}\in\mathbb{R} {^{ (M - {N_s}) \times {\widetilde N} }}$ can be given by
\begin{equation}\label{eq:12}
{\bf{\Upsilon }}{\rm{ = }}{\bf{T}}{{\bf{H}}^\dag }.
\end{equation}
The main task for offline training of ELM network is to learn the output weight ${\bf{\Upsilon }}$. With the learned output weight ${\bf{\Upsilon }}$, the chosen input weight ${\bf{W}}$ and hidden bias ${\bf{b}}$, the ELM network can implement online running, which is given in Table II.

%===画table2================
\begin{table}
	\centering
	\caption{Online procedure}
	\label{tab:table2}
	\centering
	\begin{tabular}{p{0.95\linewidth}}
	\toprule
\\
~~With the learned output weight ${\bf{\Upsilon}}$, the chosen input weight ${\bf{W}}$ and hidden bias ${\bf{b}}$, the online running steps of ELM network are summarized as follows:

\begin{enumerate}[step 1 :]
	    \item Execute the preprocessing of FS to obtain metric vector ${{\bf{\bar g }}}$, according to (\ref{eq:5})--(\ref{eq:7}).
		\item Fed ${{\bf{\bar g }}}$ into the trained ELM-based FS network to obtain network output $\bf{O}$ according to (\ref{eq:13}).
		\item From (\ref{eq:14}), estimate the frame boundary offset, i.e., obtain the estimation ${\widehat{\tau}}$.
	\end{enumerate}
	\\ \bottomrule
	\end{tabular}
\end{table}	
	
For online running, the input of ELM-based FS network (i.e., the metric vector ${{\bf{\bar g }}}$) is obtained by employing the preprocessing, i.e., the equations from (\ref{eq:5}) to (\ref{eq:7}). Then, ${{\bf{\bar g }}}$ is fed into the trained ELM-based FS network, which produces a network output ${\bf{O}} \in \mathbb{R} {^{\left( {M - {N_s}} \right) \times 1}}$ as
\begin{equation}\label{eq:13}
{\bf{O}} = \bf{\Upsilon } \cdot \sigma \left( {{\bf{W}}{\bf{\bar g }} + {\bf{b}}} \right).
\end{equation}
By expressing $\bf{O}$ as ${\bf{O}} = {\left[ {{o_{0}},{o_{1}}, \cdots ,{o_{M - {N_s} - 1}}} \right]^T}$, the estimation of frame boundary offset can be given by
\begin{equation}\label{eq:14}
\begin{aligned}
{\widehat{\tau}} = &\mathop {\arg \max }\limits_{0 \le j \le M - {N_s} - 1} {|{o_{j}}|^2}.
\end{aligned}
\end{equation}

To sum up, the ELM-based FS network is employed to improve SMs, which can overcome multi-path interfere and nonlinear distortion.

%=======================第四章 实验仿真=============================================
\section{NUMERICAL SIMULATION}

To verify the proposed ELM-based FS can improve the error probability performance, we compared it with the classical correlation-based FS \cite{c6} and the recent novel method in \cite{c15} when the nonlinear distortion is encountered. Besides, it is also necessary to validate the robustness and generalization of the performance.

%is robust and generic against varying simulation parameters.

%==========插入图2================
\begin{figure}
\centering
\includegraphics[width=0.75\columnwidth]{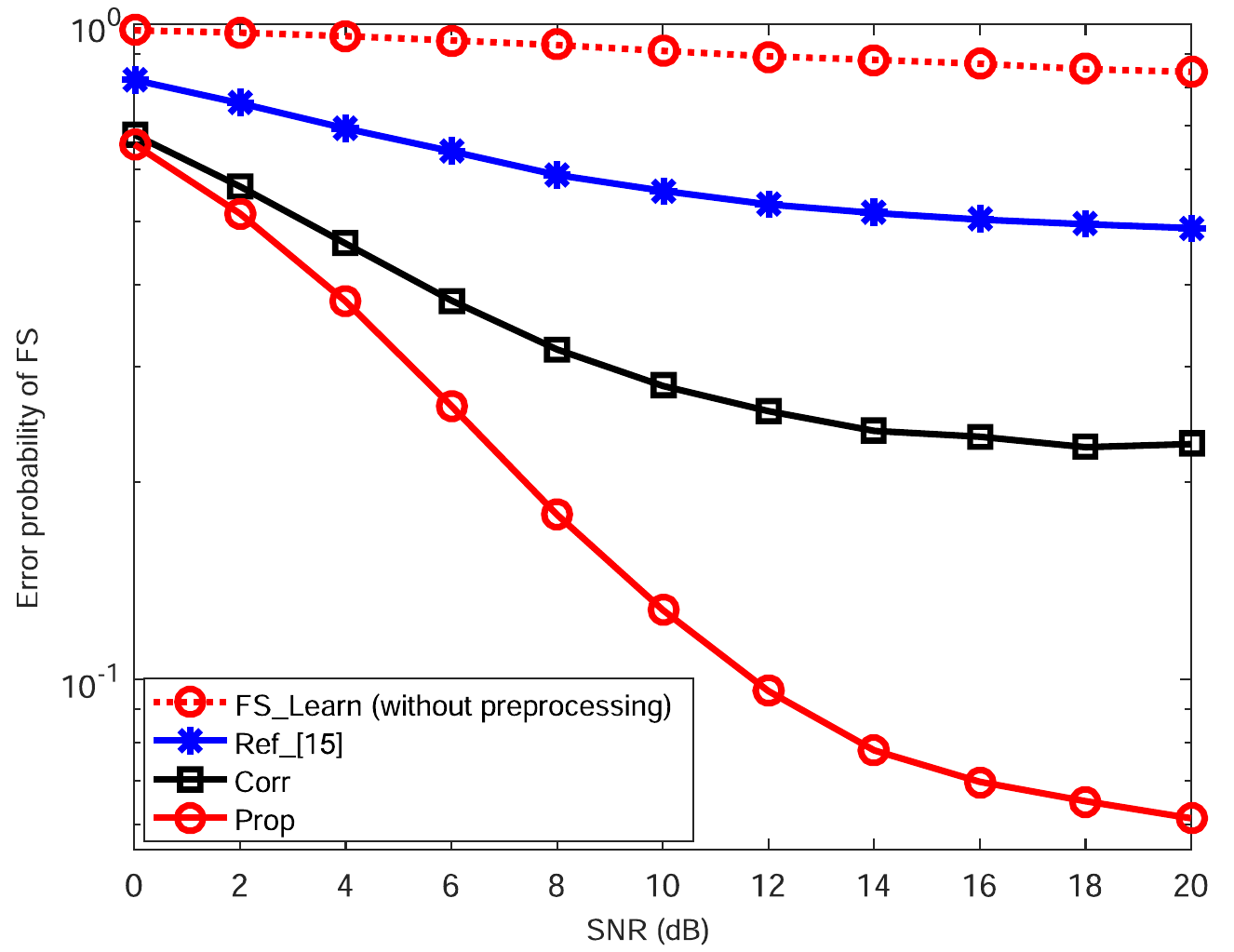}
\caption{Error probability of FS vs. SNR}
\label{fig2}
\end{figure}

\begin{figure*}[ht!]
    \centering
        \subfigure[different values of $L$]{
        \label{fig003A1}
        \includegraphics[width=0.470\columnwidth]{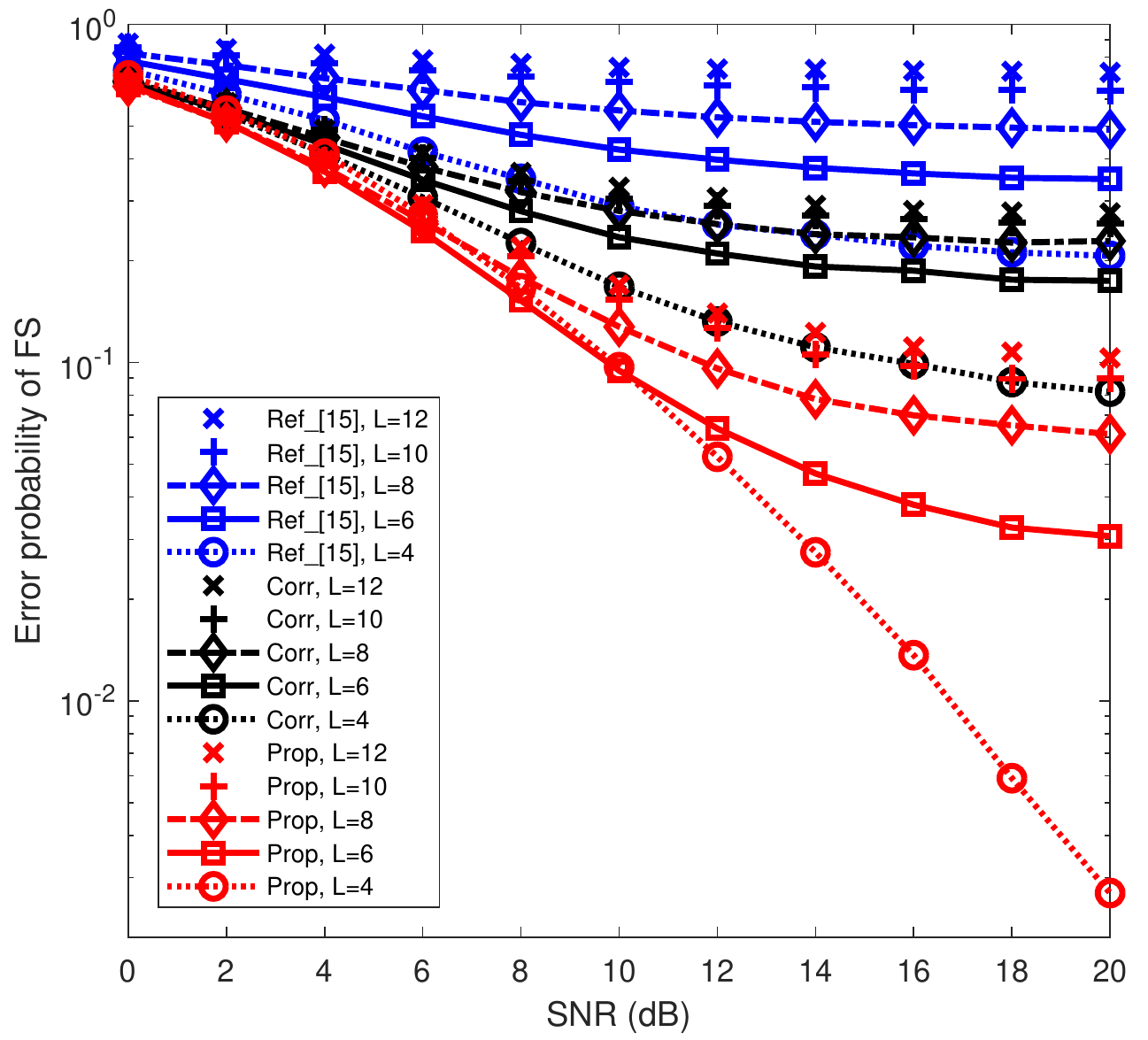}}
        \subfigure[different values of $N_s$]{
        \label{fig003A2}
        \includegraphics[width=0.470\columnwidth]{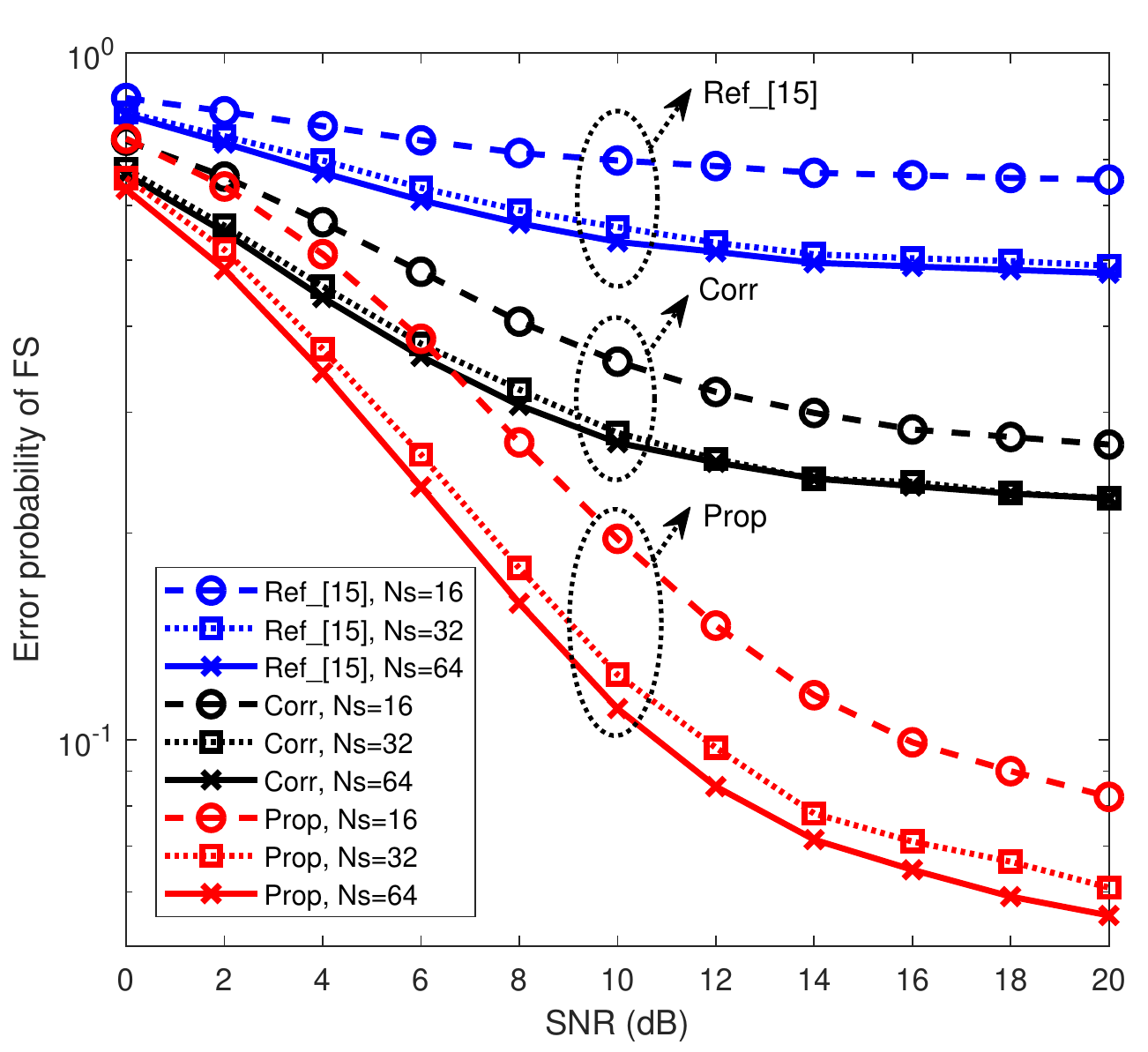}}
        \subfigure[different values of $M$]{
        \label{fig003A3}
        \includegraphics[width=0.470\columnwidth]{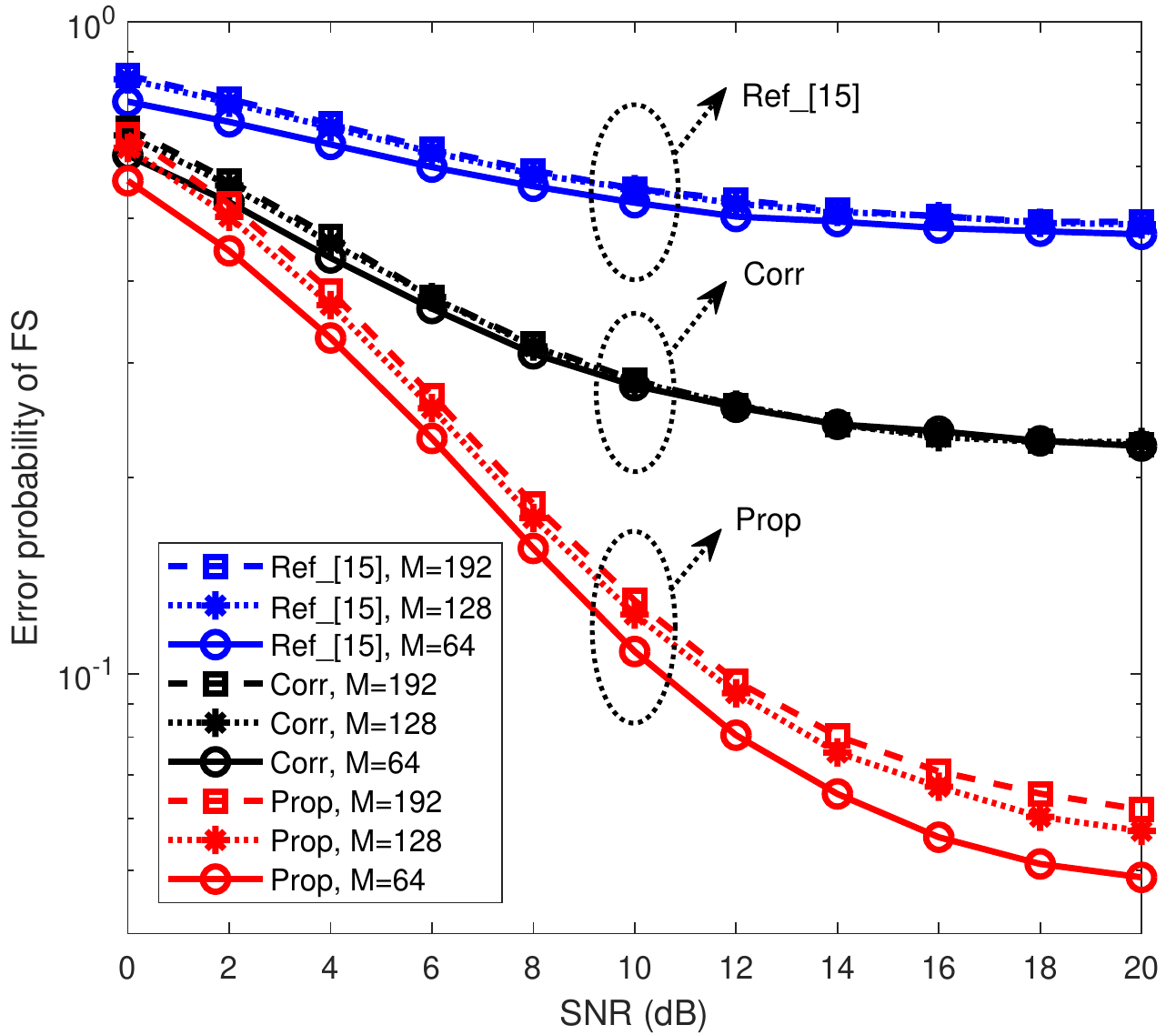}}
        \subfigure[different HPAs]{
        \label{fig003B33}
        \includegraphics[width=0.470\columnwidth]{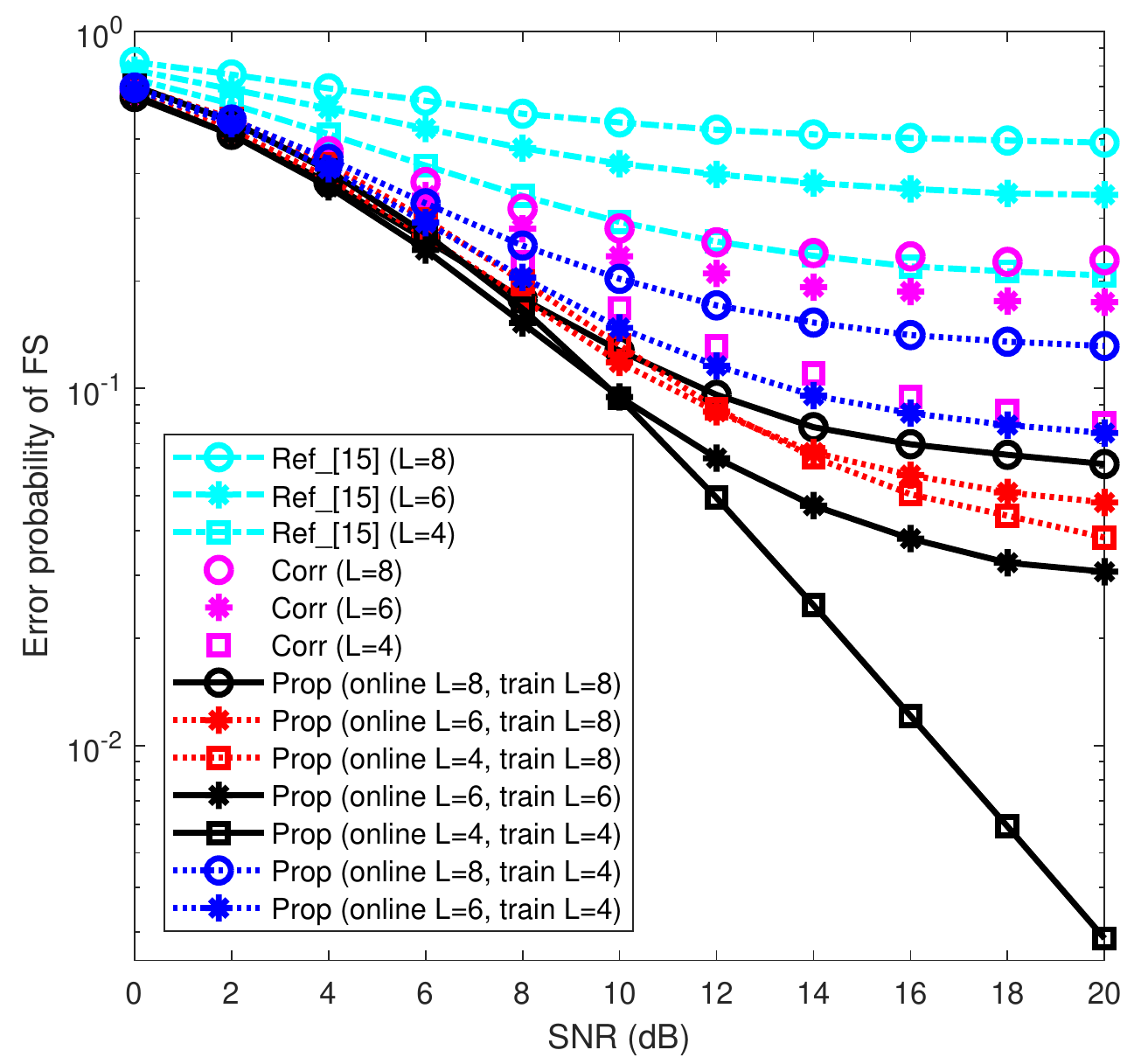}}

            \caption{Impacts of different parameters on error probability of FS}
        \label{fig3}
\end{figure*}

The basic parameters involved are listed below. The training sequence is Zadoff-Chu sequence \cite{c17}, ${N_s} = 32$, $M = 160$, $N = M-{N_s}=128$, $\widetilde N = 10N = 1280$, ${N_t} = {10^5}$, and $L = 8$ \cite{c9}. The decibel (dB) form of signal-noise-ratio (SNR) and the error probability of FS are defined as $SNR = 10{\log _{10}}\left( {P/{\sigma ^2}} \right)$ \cite{c19} and ${P_e} = {Pr} \left( {{{\widehat \tau }} \ne {\tau}} \right)$, respectively. The multi-path Rayleigh fading channel with an exponentially-decayed power coefficient (denoted as $\eta$) 0.2 is considered. For fair comparison with \cite{c15}, the same situation is considered, i.e., except the first path, each of the following $L-1$ paths is set as zero-valued with a probability of 0.5. Note that, the proposed ELM-based FS is applicable regardless of the sparsity of the channel. For nonlinear distortion, we consider the effects of HPA in this paper. The nonlinear amplitude $A\left( x \right)$ and phase $\Phi(x)$ are respectively adopted from \cite{c22}
\begin{equation}\label{eq:15}
A\left( x \right) = \frac{{{\alpha _a}x}}{{1 + {\beta _a}{x^2}}},{\rm{ }}\Phi \left( x \right) = \frac{{{\alpha _\phi }{x^2}}}{{1 + {\beta _\phi }{x^2}}}.
\end{equation}
According to \cite{c22}, ${\alpha _a} = 1.96$, ${\beta _a} = 0.99$, ${\alpha _\phi } = 2.53$, and ${\beta _\phi } = 2.82$ are considered in the simulations.

For simplicity, we use ``Prop'', ``Corr'' and ``Ref\_\cite{c15}'' to denote the proposed ELM-based FS, the correlation-based FS in \cite{c6}, and the ``CL-OMP'' FS method in \cite{c15}, respectively. In addition, ``FS\_Learn'' is used to denote the FS method that an ELM is employed to learn FS from the received observation $\bf{y}$ in (\ref{eq:1}), i.e., without the preprocessing procedure given in III-A.

\subsection{Error Probability Performance of FS}
The effectiveness of the proposed ELM-based FS is validated in terms of the error probability curves in Fig. \ref{fig2}. It could be observed that the error probabilities of ``Corr" and ``Ref\_\cite{c15}'' are much higher than that of ``Prop'' during the relatively high SNR, e.g., $SNR\geq4$dB. Meanwhile, the error probability of ``FS\_Learn'' is higher than those of ``Corr", ``Ref\_\cite{c15}'' and ``Prop''. That is, without the preprocessing procedure given in III-A, the input of ELM network is the received observation $\bf{y}$ in (\ref{eq:1}) rather than ${\bf{\bar g }}$ in (\ref{eq:7}), and thus cannot work well. It also reflects that the importance of preprocessing in ELM-based FS. In addition, the recent FS in \cite{c15} is almost not applicable due to its poor error probability even at a relatively high SNR (e.g., $Pe \geq 0.48$ at $SNR = 16$dB), while the proposed ELM-based FS achieves a relatively low error probability to retains the feasibility for practical applications for a relatively high SNR (e.g., $Pe \leq 0.06$ at $SNR = 16$dB). As a whole, the proposed ELM-based FS shows improvement of reducing error probability compared with ``Corr'' and ``Ref\_\cite{c15}''.

\subsection{Robustness Analysis}
Usually, the error probability of synchronization is influenced by the number of multi-path (i.e., $L$), the length of training sequence (i.e., $N_s$), the length of transmitted frame (i.e., $M$), and different HPAs (i.e., different values of nonlinear distortion). To illuminate the robustness of improvement under nonlinear distortion, Fig. \ref{fig3}(a), Fig. \ref{fig3}(b), Fig. \ref{fig3}(c) and Fig. \ref{fig3}(d) are given to demonstrate the impacts against $L$, $N_s$, $M$, and different HPAs, respectively. Except for the change of the impact parameters (i.e., only $L$, $N_s$, $M$ and the parameters of HPA are changed for Fig. \ref{fig3}(a), Fig. \ref{fig3}(b), Fig. \ref{fig3}(c) and Fig. \ref{fig3}(d), respectively), other basic parameters remain the same as Fig. \ref{fig2} during the simulations.

\subsubsection{Robustness against $L$}
To demonstrate the impact of $L$ on robustness, Fig. \ref{fig3}(a) shows the error probability of FS, where $L=4$, $L=6$ and $L=8$ are considered. It is observed from Fig. \ref{fig3}(a), the improvement of reducing error probability is more significant with a smaller $L$. With the increase of $L$, the error probabilities increase for all cases (i.e., ``Corr'', ``Ref\_\cite{c15}'' and ``Prop''), due to the stronger multi-path interference. Even so, the error probability of ``Prop'' is much lower than those of ``Corr'' and ``Ref\_\cite{c15}'', especially for $SNR \geq 12$dB. As a result, compared with those of ``Corr'' and ``Ref\_\cite{c15}'', the proposed ELM-based FS exhibits the robust improvement of reducing error probability against varying $L$.

\subsubsection{Robustness against $N_s$} Fig. \ref{fig3}(b) plots the error probability of FS with different $N_s$ (i.e., $N_s=16$, $N_s=32$ and $N_s=64$). From Fig. \ref{fig3}(b), a lower error probability of FS can be obtained as $N_s$ increases for the cases of ``Corr'', ``Ref\_\cite{c15}'' and ``Prop''. This is because a longer training sequence is more effective for overcoming multi-path interference in the given scenario $L=8$. The error probability of ``Prop'' is lower than those of ``Corr'' and ``Ref\_\cite{c15}'', especially for the high SNR regime (e.g., $SNR \geq 14$dB). By utilizing the proposed ELM-based FS, the error probability is lower than those of ``Corr'' and ``Ref\_\cite{c15}'', and the change of $N_s$ shows less impact on this improvement.

\subsubsection{Robustness against $M$}
To validate the effectiveness against the impact of $M$, the error probability curves are illustrated in Fig. \ref{fig3}(c), where $M=192$, $M=128$ and $M=64$ are considered, respectively. As $M$ decreases, the error probabilities of ``Corr'', ``Ref\_\cite{c15}'' and ``Prop'' slightly decrease due to the reduced locations for index search (since $0\leq \tau \leq M-N_s-1$). From Fig. \ref{fig3}(c), the error probability of ``Prop'' is lower than those of ``Corr'' and ``Ref\_\cite{c15}'' given different values of $M$. This reflects that the error probability is reduced and the improvement is robust against varying $M$.

\subsubsection{Robustness against different HPAs}

Besides the HPA mentioned above (denoted as HPA1), an additional HPA (denoted as HPA2), which parameters are set as ${\alpha _a} = 1.66$, ${\beta _a} = 0.06$, ${\alpha _\phi } = 0.15$, and ${\beta _\phi } = 0.35$ \cite{c22}, is also employed in Fig. \ref{fig3}(d) to observe the influence of HPA on ``Prop''. From \cite{c22}, the root mean-square (RMS) errors of nonlinear amplitude and phase of HPA1 (HPA2) are 0.012 (0.041) and 0.478 (0.508), respectively. According to the RMS errors, HPA1 has less distortion than HPA2, and thus brings ``Prop'', ``Corr'' and ``Ref\_\cite{c15}'' lower error probability of FS. Especially, for both HPA1 and HPA2, the error probability of ``Prop'' is obviously lower than those of ``Corr'' and ``Ref\_\cite{c15}''. Therefore, the proposed ELM-based FS can work well with HPA1 and HPA2.

\subsection{Generalization Analysis}
Fig. \ref{fig4}(a) and Fig. \ref{fig4}(b) present the generalization performance against $L$ and $\eta$ (i.e., the decayed power coefficient), respectively.

%illustrates the impact of multi-path (i.e., $L$), the exponentially-decayed power coefficient (i.e., $\eta$) and the HPA distorted values to generalization performance, respectively.
\subsubsection{Generalization against $L$}

 In Fig. \ref{fig4}(a), the trained networks of $L=4$ and $L=8$ are respectively employed to test the cases where $L=4$, $L=6$, and $L=8$. From Fig. \ref{fig4}(a), the performance of error probability is degraded when the testing $L$ is not the training $L$. Even so, the error probability of ``Prop'' is obviously lower than those of ``Corr'' and ``Ref\_\cite{c15}''. Therefore, for the cases where testing $L$ is not training $L$, the ``Prop'' still improves the error probabilities of ``Corr'' and ``Ref\_\cite{c15}''.

%
%There are some curves to verify the generalization of $L$ in Fig. \ref{fig4}(a). It could be observed that whether online $L=6,8$, training $L=4$ or online $L=4,6$, training $L=8$, $L$ influences the generalization performance in reducing error probability to some extent, but compared with ``Ref\_\cite{c15}'' and ``Corr'', the ``Prop'' still can improve the performance of error probability.

\subsubsection{Generalization against $\eta$}

\begin{figure}
    \centering
        \subfigure[against $L$]{
        \label{fig003B1}
        \includegraphics[width=0.470\columnwidth]{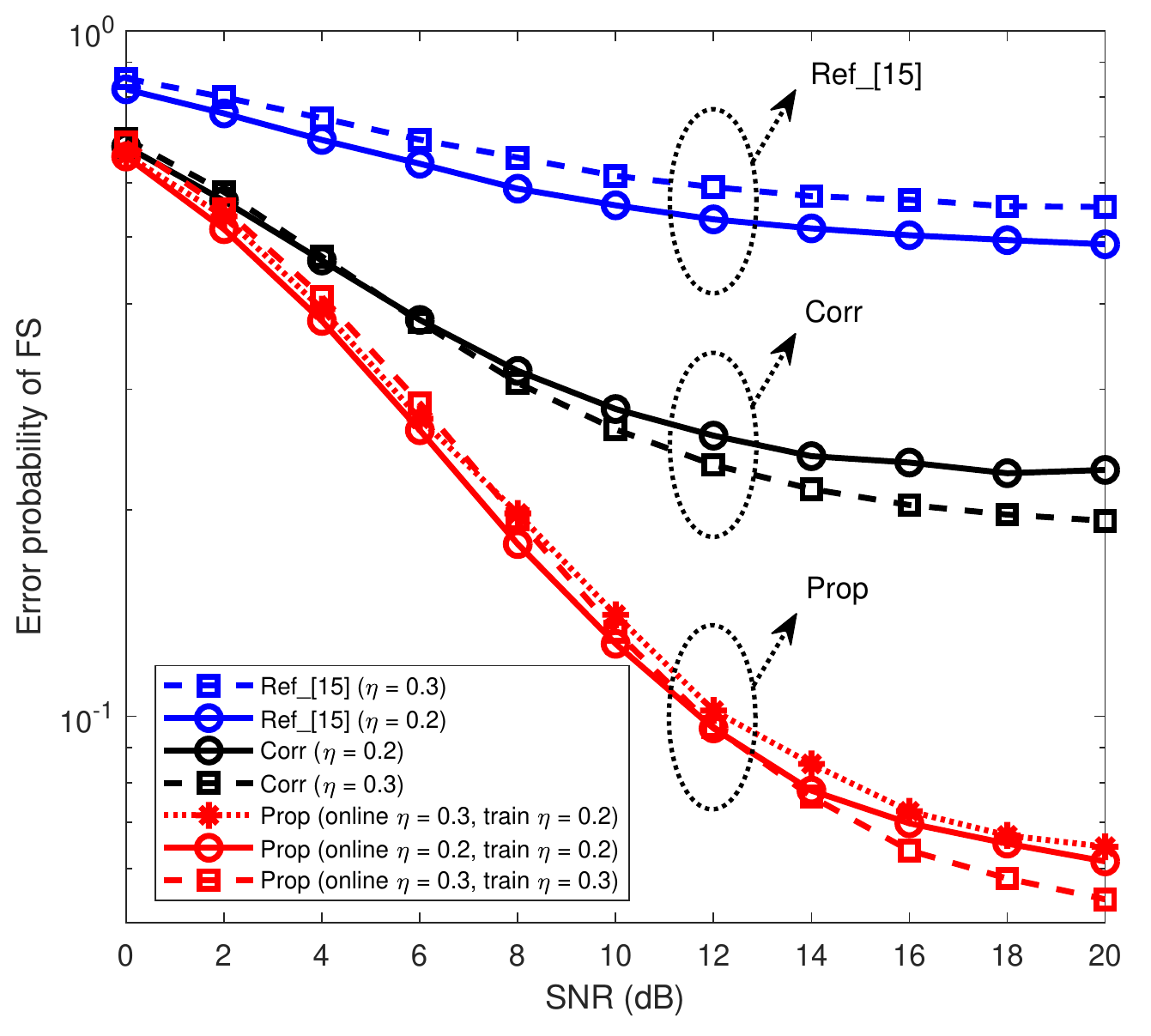}}
        \subfigure[against $\eta$]{
        \label{fig003B2}
        \includegraphics[width=0.470\columnwidth]{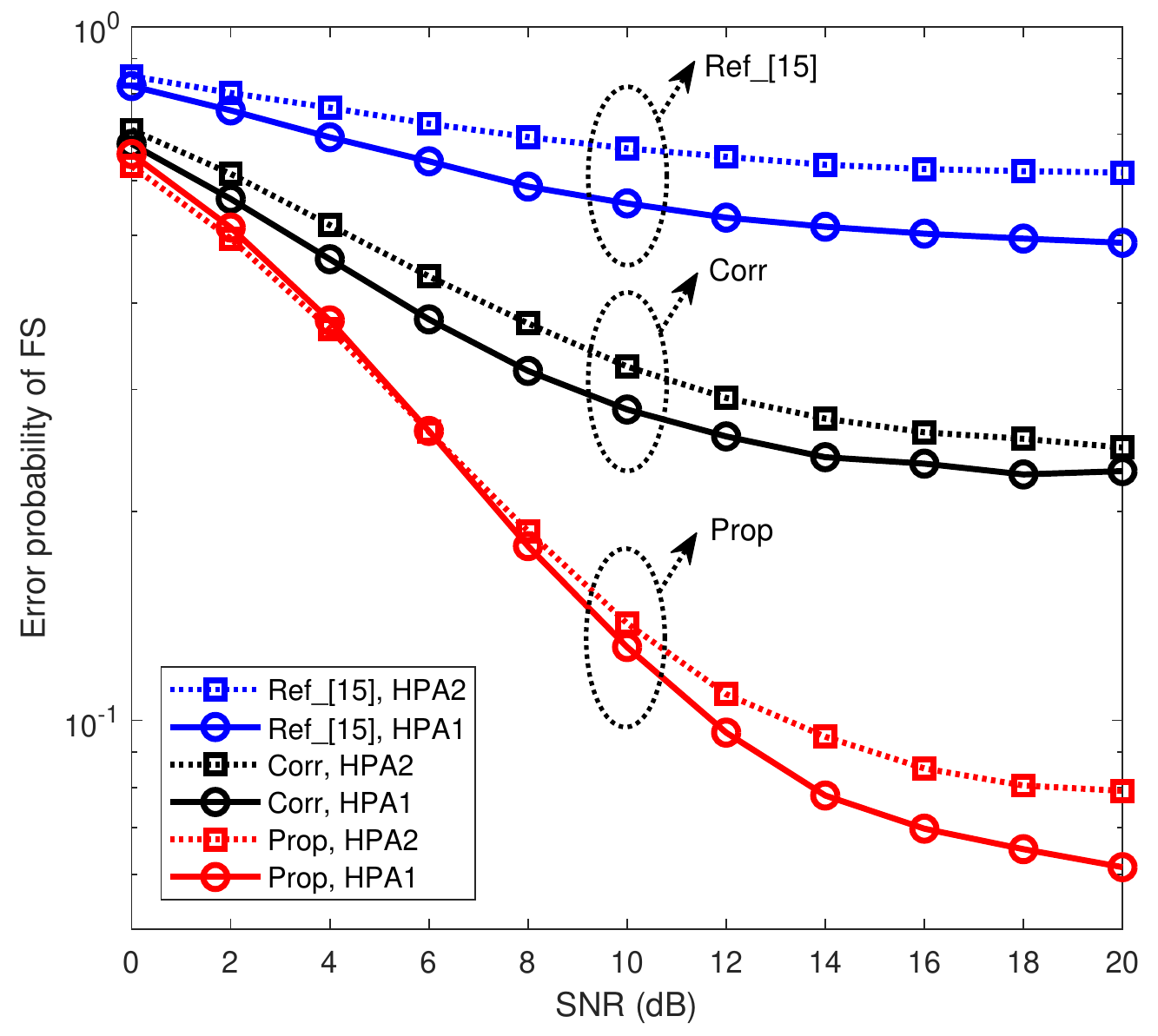}}
           \caption{Generalization analysis against $L$ and $\eta$}
    \label{fig4}
\end{figure}

The error probability performance for the case where the testing $\eta$ is not the training $\eta$ as plotted in Fig. \ref{fig4}(b). In Fig. \ref{fig4}(b), the training $\eta$ is 0.2, while the testing $\eta$ is 0.3. According to the error probability of FS, this influence is not obvious for ``Prop''. Besides, the error probability of ``Prop'' is obviously lower than those of ``Corr'' and ``Ref\_\cite{c15}''. Thus, the ``Prop'' possesses a good generalization performance against $\eta$.

%--------------------第五章 结论-------------------------------------------
\section{CONCLUSION}
 In this work, we investigated the ELM-based FS to improve the performance of burst-mode communication systems. A preprocessing is first performed to capture the coarse features of SM, followed by an ELM network to reduce system's nonlinear distortion and recover SMs. Compared with the existing methods, the proposed ELM-based FS is validated with its robustness and generalization by reducing error probability. In this paper, the difficulty of obtaining desired labels is simplified by generating them according to the existing channel model. In our future works, we will consider the desired FS labels in real channel scenarios to promote the application of machine learning-based FS in practical systems (such as IoT, WLAN, etc) with nonlinear-distortion.

% The proposed ELM-based FS can be applied in many systems, such as IoT, WLAN, etc., and can work for the scenarios with nonlinear distortion. Our future work is to investigate effective machine learning-based FS methods,

 %We also find that the robustness and generalization ability with ELM-based FS method still limited. Therefore, how to make the channel model closer to the real wireless channels will be considered in the future. We will make more deep analysis and complex calculations for the future work.

% that's all folks
\end{document}